# Asymmetric Acoustic Energy Transport in Non-Hermitian Metamaterials

Ramathasan Thevamaran[1,=,*], Richard Massey Branscomb[2,=], Eleana Makri[3], Paul Anzel[4],

Demetrios Christodoulides[5], Tsampikos Kottos[3], Edwin L. Thomas[2,*]

[1]*Department of Engineering Physics, University of Wisconsin-Madison, Madison, WI 53706.*

[2]*Department of Materials Science and NanoEngineering, Rice University, Houston, TX 77005.*

[3]*Department of Physics, Wesleyan University, Middletown, CT 06459.*

[4]*HEB, San Antonio, TX 78204.*

[5]*Center for Research and Education in Optics and Lasers (CREOL), College of Optics and Photonics,*

*University of Central Florida, Orlando, FL 32816.*

[*]Corresponding Authors: RT (thevamaran@wisc.edu); ELT (elt@rice.edu)

[=]These authors contributed equally to this work.

## Abstract

The ability to control and direct acoustic energy is essential for many engineering applications such as vibration and noise control, invisibility cloaking, acoustic sensing, energy harvesting, and phononic switching and rectification. The realization of acoustic regulators requires overcoming fundamental challenges inherent to the time-reversal nature of wave equations. Typically, this is achieved by utilizing either a parameter that is odd-symmetric under time-reversal or by introducing passive nonlinearities. The former approach is power consuming while the latter has two major deficiencies: it has high insertion losses and the outgoing signal is harvested in a different frequency than that of the incident wave due to harmonic generation. Here, we adopt a unique approach that exploits spatially distributed linear and nonlinear losses in a fork-shaped resonant metamaterial. Our compact design demonstrates asymmetric acoustic reflectance and transmittance, and acoustic switching. In contrast to previous studies, our non-Hermitian metamaterial exhibits asymmetric transport with high frequency purity of the outgoing signal.

Controlling wave propagation using metamaterials—composite materials with a specifically tailored impedance profile—has led to many unprecedented technologies in recent years. Photonic (Joannopoulos et al., 2008; Saleh and Teich, 1991) and phononic (Deymier, 2013) crystals, negative and modulated index materials (Eleftheriades and Balmain, 2005; Lakes, 2001; Nassar et al., 2017), cloaking (Werner and Kwon, 2014) and super-resolution systems (Zalevsky and Mendlovic, 2004) are testaments to the increasing ability to design and fabricate dielectric and mechanical impedance profiles of the underlying structures that can modify the propagation of light and sound to yield exotic wave phenomena. Prospects of utilizing loss or absorption as a critical ingredient in metamaterials is emerging as a new design paradigm (El-Ganainy et al., 2018; Hodaei et al., 2017; Konotop et al., 2016; Kottos and Aceves, 2016; Suchkov et al., 2016). Tailoring loss, together with a judicious design of the impedance profile, can lead to the realization of devices with unconventional properties and novel functionalities. The best-known example of such non-Hermitian methodologies is the creation of Parity-Time (PT) symmetric materials. These materials steer light (El-Ganainy et al., 2007; Makris et al., 2008; Musslimani et al., 2008) and sound (Fleury et al., 2015a; Shi et al., 2016; Zhu et al., 2014) propagation using loss and/or gain, in stark contrast to the previous approaches in photonic and phononic devices, (Boechler et al., 2011; Liang et al., 2010) and metamaterials (Haberman and Guild, 2016; Lee et al., 2012; Milton et al., 2006) that primarily employ differences in the real part of the dielectric or mechanical impedance of their constituents or purposefully insert defect modes within a bandgap. Novel technologies that can emerge from the manipulation of loss and gain include shadow-free sensing (Fleury et al., 2015a), unidirectional transparency (Zhu et al., 2014), coherent perfect absorption (Merkel et al., 2015; Song et al., 2014), and asymmetric transmission (Popa and Cummer, 2014). Active elements are incorporated through piezoelectric and piezoacoustic effects in those systems to provide gain and loss. For instance, speakers (Fleury et al., 2015a; Shi et al., 2016) and nonlinear electronic circuits (Popa and Cummer, 2014) have been used to control the flow of sound. Nevertheless, such elements face challenges in terms of energy consumption and physical size. Passive conservative nonlinearities have also been used for achieving asymmetric transmission (Boechler et al., 2011; Lepri and Casati, 2011; Lepri and Pikovsky, 2014). In many of these cases, the output signal is harvested at a different frequency than the input signal in such conservative nonlinear systems.

Here, we demonstrate asymmetric acoustic reflection in a linear non-Hermitian metamaterial (LnH-MetaMater), and asymmetric acoustic reflection and transmission in a nonlinear non-Hermitian metamaterial (nLnH-MetaMater). In contrast to the majority of existing nonlinear acoustic rectifiers (see recent reviews (Fleury et al., 2015b; Maznev et al., 2013)), our design provides a compact subwavelength system with high frequency purity that utilizes the natural losses of the constituent materials without using any external power.

**Non-Hermitian metamaterial design.** To realize asymmetric acoustic energy transport exploiting non-Hermiticity, we designed a metamaterial such that the two resonant components of the metamaterial have dissimilar imaginary parts of the complex resonance frequency while the real parts remain approximately equal. The metamaterial consists of two air-coupled tuning forks that resonate at a desired matched frequency: a metal fork made of aluminum (Al) (density $\rho$=2700 kg m$^{-3}$, Young's modulus E=69 GPa, Poisson's ratio $\nu$=0.33) and a polymer fork made of polycarbonate (PC) ($\rho$=1200 kg m$^{-3}$, E=3 GPa, $\nu$=0.4). We modify their resonance mode of interest to be at the same frequency by tailoring the geometry of each fork. We utilize both the principal and the sway modes of resonance and fabricate two pairs of forks that are frequency matched at their respective modes. At the matched frequency of resonance, the Al fork acts as the low-loss element while the PC fork acts as the high-loss element. An intriguing feature of this design approach is that it permits constituents with different material properties and geometries provided the resonators resonate at the same frequency. It allows designing metamaterials with dissimilar imaginary parts without affecting the ability to maintain similar real parts.

A tuning fork has two primary modes of resonance (Rossing, 1992): (i) the principal mode at which the tines of the forks move out-of-phase (Fig.1 a,b), and (ii) the sway mode at which the tines of the forks move in-phase (Fig.1 c,d). We fabricated two forks from each material employing different dimensions to have their respective resonant frequencies matched at approximately 2175 Hz for principal mode (corresponds to a wavelength in air of 159 mm) and at approximately 1925 Hz for the sway mode (corresponds to a wavelength in air of 175 mm). The principal mode is less susceptible to the fork stem mounting boundary conditions as the oscillation of the tines occurs about a node at the top of the stem (node 1 in Fig.1 a,b). The response of both the Al and the PC forks at their principal mode is linear and independent of the excitation amplitude for the range of incident sound pressure amplitudes we use in this study (node 2 in Fig.1 a,b; Fig.S2) while exciting the forks at the sway mode induces strong nonlinear behavior in the PC fork (Fig.1 c,d; Fig.S3). We characterized the loss behavior of each fork by its full-width-at-half-maximum (FWHM) of the resonance. The measured loss (FWHM) at the principal mode is 5.60±0.40 Hz for the Al fork, and 22.08±0.84 Hz for the PC fork for all excitation amplitudes (Fig.1 a,b). Our experiments also indicate that the transmittance and reflectance are not equal for Al and PC forks (Fig.S2).

In contrast to the principal mode, the sway mode oscillation is about a node at the bottom of the stem and it strongly couples to the system boundary through the fork's stem—a mechanism that leads to a strongly nonlinear response (Fig.1 c,d) at the sway-mode resonance. The strength of this nonlinear response is typically determined by the material properties. Its nature—dissipative or conservative—can be quantified by a detailed analysis of the variations of the position of the resonance peak (for conservative) and of the FWHM of the resonance peak $\gamma$ (for dissipative) as a function of the excitation amplitude. In the Al fork for example, we find that the FWHM of its sway-mode resonance is minimally affected by the excitation amplitude (Fig.1e). We, therefore, consider the resonance peak of the Al fork (Fig.1e) to be affected only by conservative nonlinearities present within the range of excitation amplitudes used in our experiments.

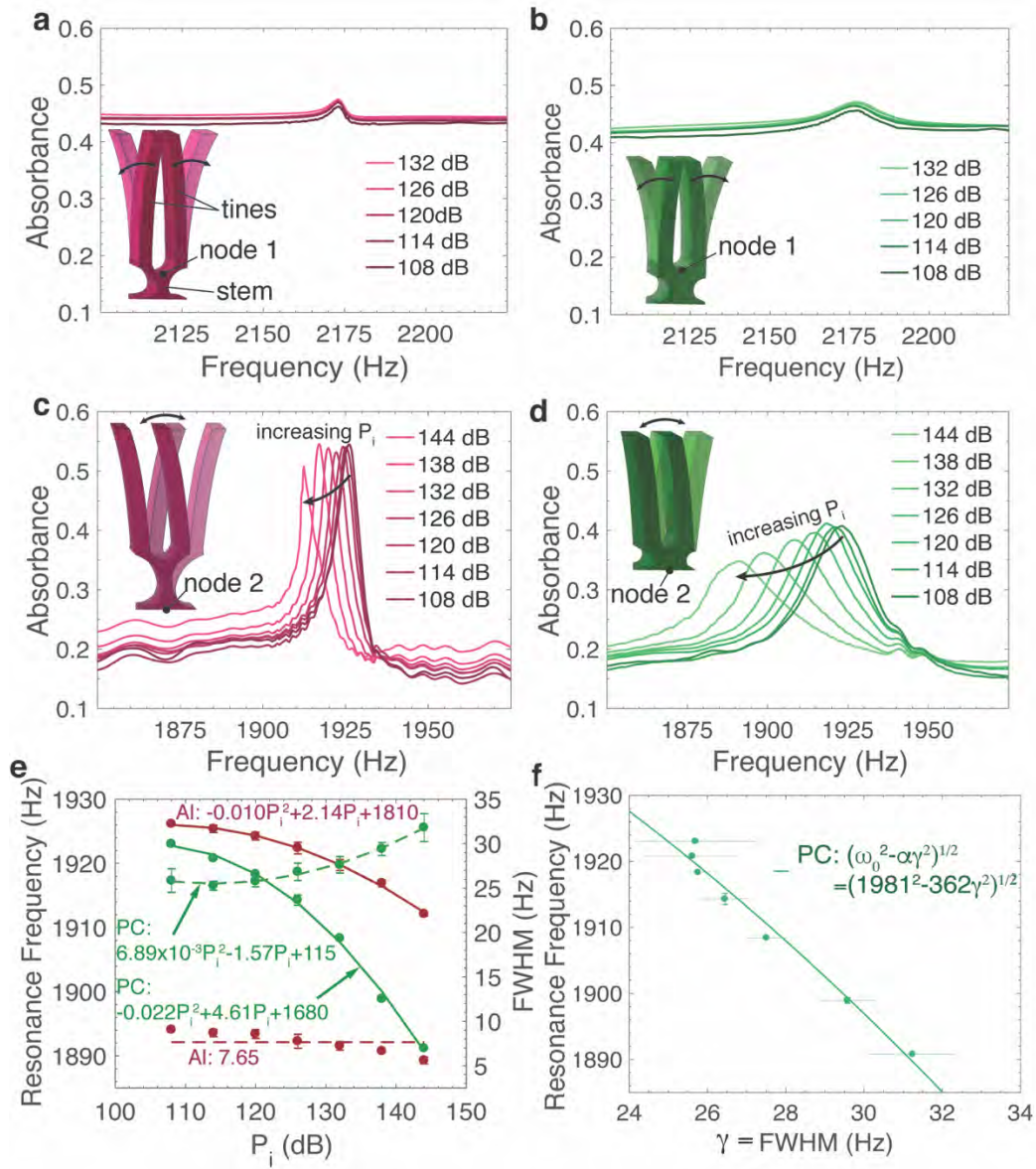

**Figure 1| Resonant behavior of the tuning forks at their principal and sway modes. a, b,** absorbance at the individual resonances of an Al (a) and a PC (b) forks at their principal mode (~2175 Hz); insets show the schematic illustration of the principal mode displacement of the respective tuning forks. **c, d,** absorbance at the individual resonances of an Al (c) and a PC (d) fork at their sway mode (~1925 Hz); insets show the schematic illustration of the sway mode displacement of the respective tuning forks. **e,** dependence of the peak frequencies (solid lines) and the losses (FWHM) (dashed lines) of the sway mode of resonance on the excitation amplitude ($P_i$). **f,** dependence of resonance frequency of the PC fork on its loss, γ (FWHM).

The PC fork, however, shows a strong increase of the FWHM of the sway mode as a function of the excitation amplitude, $P_{in}$ (Fig.1e)—a clear indication that the PC fork experiences nonlinear losses. The observed shift in the resonance peak (Fig.1e) is primarily a direct consequence of these nonlinear losses. This dependence of resonance peak frequency ($\omega$) on the loss ($\gamma$) is modeled as $\omega = \sqrt{\omega_0^2 - \alpha\gamma^2}$, where $\omega_0$ is the linear frequency of the PC fork, and $\alpha$ is a fitting parameter (Fig.1f). In the case of the PC fork, the losses are nonlinear and this is also reflected in the resonance shift and peak broadening. The agreement of the experimentally observed response with the above damped-oscillator model, allows us to exclude the existence of any other strong conservative nonlinear mechanisms in the resonance of the PC fork.

Our LnH-MetaMater consists of an Al fork and a PC fork that have the same resonance frequency at their principal mode, and our nLnH-MetaMater metamaterial consists of a different pair of Al and PC forks designed to have their resonance frequency matched at the sway mode. The forks are placed at a designated gap, G, thereby forming a resonance cavity between them. We devised a two-port acoustic testing setup (Åbom, 1991; *ISO 10534-2:1998(E)*) to experimentally investigate the acoustic transport in the LnH-MetaMater and the nLnH-MetaMater (Fig.2, Fig.S1). The forks are placed inside an impedance tube where the plane wave from a speaker is incident on the metamaterial in port-A. The incident wave is partially transmitted into port-B while the rest is absorbed and/or reflected back into port-A. Back reflections from the end of port-B are minimized by an absorbent foam. Two pressure-field impedance-tube microphones in each port are used to measure the incident and reflected sound waves. We implemented a lock-in amplification system (Molerón et al., 2015) for phase-sensitive measurements of the waves. From the measured pressure, we calculated the reflectance ($R = \frac{P_r P_r^*}{P_i P_i^*}$) and transmittance ($T = \frac{P_t P_t^*}{P_i P_i^*}$), where $P_i$ is the incident pressure wave, $P_r$ is the reflected pressure wave, and $P_t$ is the transmitted pressure wave, see methods section in supplementary information. We used two sample configurations of the metamaterial as shown in Fig.2 to investigate the directional wave propagation characteristics. Since the structure incorporates losses, the sound flux is not conserved, and the absorbance A is defined as $A \equiv 1 - T - R$.

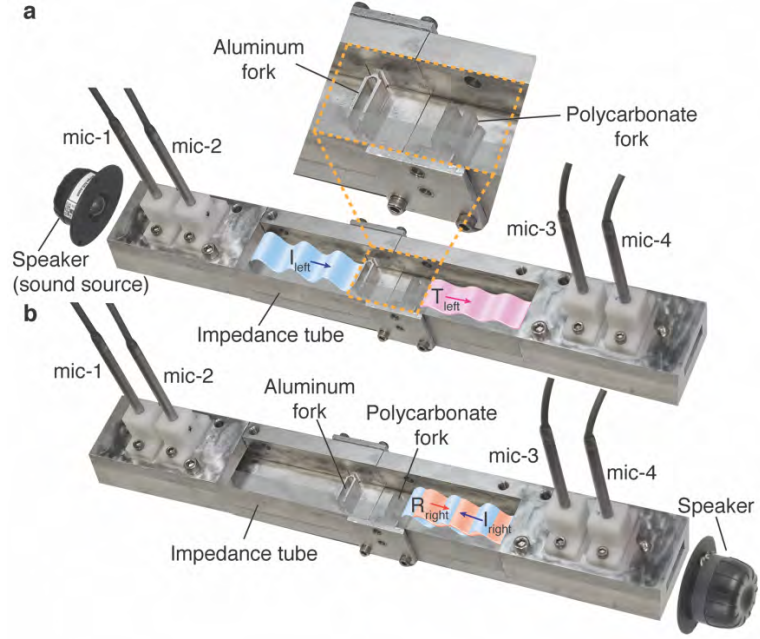

**Figure 2| Experimental apparatus used for measuring the asymmetric acoustic energy transport.** The metamaterial is embedded in a two-port acoustic testing system for phase-sensitive measurement of the incidence, reflection, and transmission: **a,** the wave incident from the left (incident first on Al fork) (inset shows a magnified view of the forks embedded in the waveguide). **b,** the wave incident from the right (incident first on PC fork).

**The linear non-Hermitian metamaterial (LnH-MetaMater).** To investigate the asymmetric wave propagation in the LnH-MetaMater, we plot the degree of asymmetry in reflectance, transmittance, and absorbance characterized by $Q_R = \frac{R_{PM} - R_{MP}}{R_{PM} + R_{MP}}$, $Q_T = \frac{T_{PM} - T_{MP}}{T_{PM} + T_{MP}}$, and $Q_A = \frac{A_{PM} - A_{MP}}{A_{PM} + A_{MP}}$ as functions of the frequency, and the incident sound pressure level ($P_i$) (Fig.3). Here, the subscript $MP$ of $R$, $T$, and $A$ denotes that the wave is first incident on the metal (Al) fork (left incidence), and the subscript $PM$ denotes the wave is first incident on the polymer (PC) fork (right incidence). The color density quantitatively describes the $Q_R$, $Q_T$, and $Q_A$ in each plot. The two forks in the LnH-MetaMater have same design frequency of ~2175 Hz, but different losses at the principal mode of resonance (Fig.1 a,b). The reflectance (Fig.3c) exhibits directional asymmetry at the resonance frequency while the transmittance (Fig.3d) remains nearly symmetric. The color on the positive scale of $Q_R$ implies that the intensity of the reflectance corresponding to the right incidence is higher than that of the left incidence (also see Fig.S4).

The asymmetric reflectance is a direct consequence of non-uniform losses present in the metamaterial and their influence at various reflection paths that contribute to the left and right total reflection. Incident sound waves from the right (PC) side will have a total reflection that can be written in the form of a geometric series $r_{tot}^{PM} = r^{PC} + t^{PC} r^{Al} t^{PC} + t^{PC} r^{Al} r^{PC} r^{Al} t^{PC} + \cdots = r^{PC} + t^{PC} r^{Al} \frac{1}{1 - r^{PC} r^{Al}} t^{PC}$, where $t^{PC/Al}$ and $r^{PC/Al}$ are the transmission and reflection amplitudes of the PC/Al forks respectively. Similarly, the total reflection for incident waves encountering the metamaterial from the left (Al) side can be written as $r_{tot}^{MP} = r^{Al} + t^{Al} r^{PC} t^{Al} + t^{Al} r^{PC} r^{Al} r^{PC} t^{Al} + \cdots = r^{Al} + t^{Al} r^{PC} \frac{1}{1 - r^{PC} r^{Al}} t^{Al}$. The convergence of the geometric series is guaranteed since $|t^{PC}|^2, |r^{PC}|^2, |t^{Al}|^2, |r^{Al}|^2 \ll 1$ due to absorption (Fig.1 a,b and Fig.S2). Direct comparison of $r_{tot}^{MP}$ with $r_{tot}^{PM}$ indicates that these two expressions are not the same. Their difference is already evident at the first term $r_{tot}^{PM/MP} \approx r^{PC/Al}$, which represents wave-paths associated with direct reflection from the PC/Al fork.

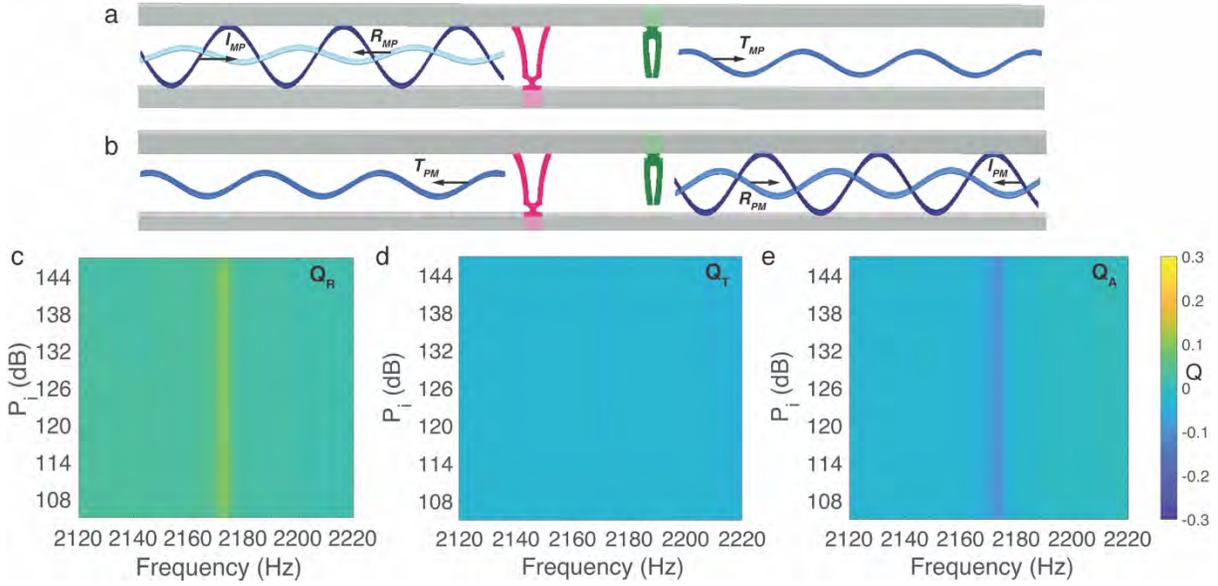

**Figure 3| Experimental response of the LnH-MetaMater. a,b,** Illustrations of the asymmetric response of the LnH-MetaMater for left and right incidences. Asymmetry in **c,** reflectance ($Q_R$), **d,** transmittance ($Q_T$), and **e,** absorbance ($Q_A$) as functions of frequency and incident sound pressure amplitude $P_i$ for a gap between the forks, G=157 mm (~$\lambda$).

At the principal-mode resonance frequency, we observe that the $Q_R$>0 (Fig.3 c) and the $Q_A$<0 (Fig.3 e) indicating that the reflection is higher, and absorption is lower when the sound is incident from right (PC fork) side. Since the PC fork is a strong-loss element, the incident waves from the right side gets highly reflected immediately. This is a manifestation of an overdamping behavior. However, the waves incident on the LnH-MetaMater from the Al side (left incidence) are transmitted into the resonant cavity formed by the two forks and dwell in the cavity due to multiple scattering events between the two forks (see the various terms in the geometric series above). These multiple scattering events with the PC fork lead to an overall enhancement of the absorption and consequently decrease the overall reflection, thus resulting in asymmetric reflection (and absorption) by the LnH-MetaMater. We, therefore, have an asymmetric resonance-enhanced absorption due to the non-uniform losses in the metamaterial.

Although the reflectance and absorbance in the LnH-MetaMater are asymmetric, the transmittance is symmetric (Fig.3d). Similar considerations as those for the reflectance indicate that the transmission paths for both left and right incidences contain the exact same scattering events resulting in symmetric transmission. Additionally, it is evident from these color density plots that the response of the LnH-MetaMater does not show any dependence on the amplitude of the incident pressure wave, as expected for a linear system.

**The nonlinear non-Hermitian metamaterial (nLnH-MetaMater).** The nLnH-MetaMater, on the other hand, consists of an Al fork and a PC fork having closely matched frequencies (~1925 Hz), but very dissimilar losses at the sway mode of resonance (Fig.1c,d) with the PC fork exhibiting predominantly dissipative nonlinearities (Fig.1e,f). The nLnH-MetaMater exhibits left and right asymmetry in transmittance (Fig. 4d) in addition to asymmetric reflectance (Fig.4c) and absorbance (Fig.4e). The transmission asymmetry is most pronounced in the regions near resonance, and is predominantly a result of the dissipative nonlinearity (i.e. nonlinear imaginary part of the resonance) present in the PC fork (the contrast ratio of transmission asymmetry ($T_{MP}/T_{PM}$), which reaches values above 5 within the range of amplitudes tested, can be seen in Fig.S7). Because we utilize dissipative nonlinearity, the outgoing signal has high frequency purity unlike in the use of conservative nonlinearities (Boechler et al., 2011; Lepri and Casati, 2011; Lepri and Pikovsky, 2014)—i.e. there are no significant higher harmonics observed, for example, as seen in the frequency response of the nLnH-MetaMater for the incidence at 1903 Hz and 1919 Hz (Fig.4f,g). These two incidence frequencies correspond to the highest asymmetry in transmission (Fig.4d). As shown in the frequency response measured in the empty impedance tube with no forks, (Fig.4h), the higher harmonics seen in Fig.4f and g have been generated by the speaker.

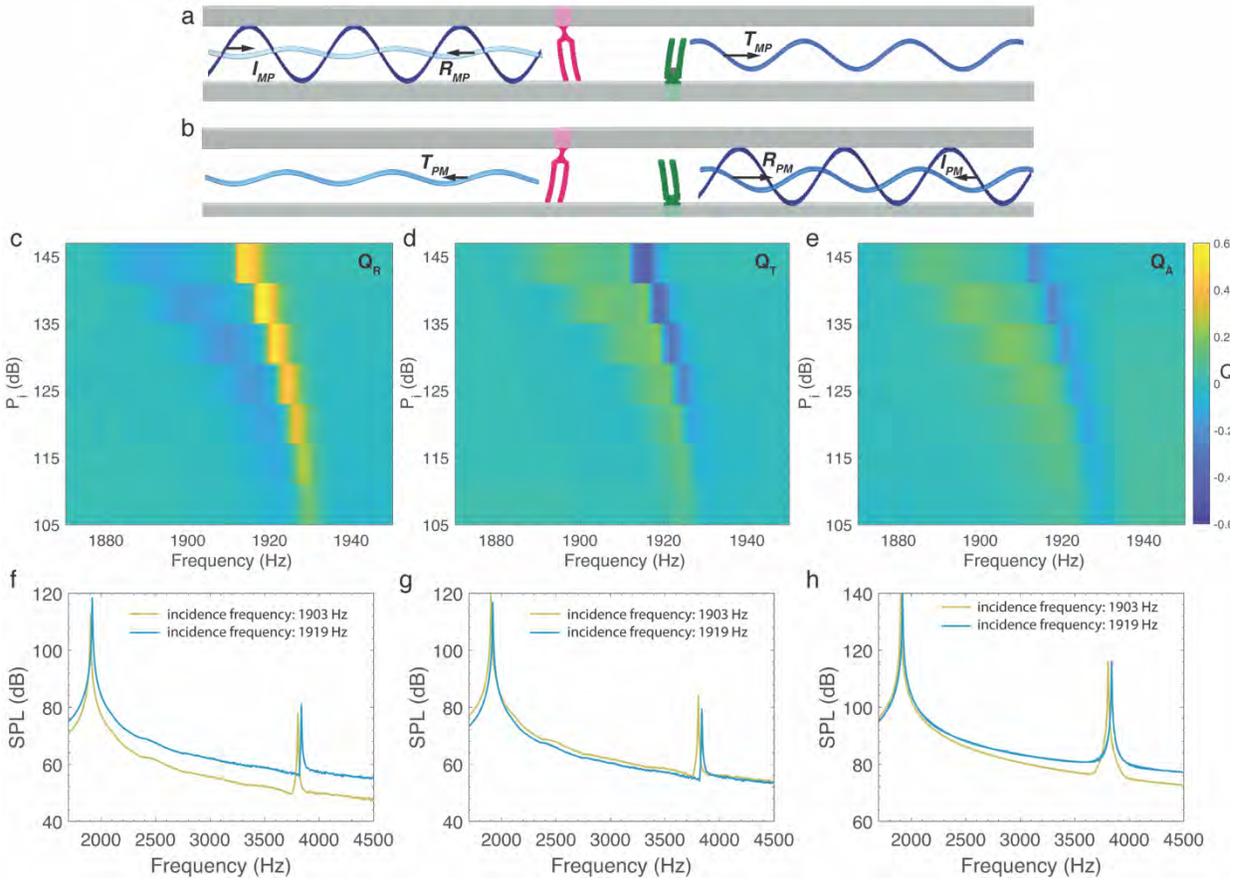

**Figure 4| Experimental response of the nLnH-MetaMater. a,b,** Illustrations of the asymmetric response of nLnH-MetaMater for left and right incidences as a function of frequency and incident sound pressure amplitude $P_i$ at the gap between the forks G=157 mm (~$\lambda$). Asymmetry in **c,** reflectance ($Q_R$), **d,** transmittance ($Q_T$), and **e,** absorbance ($Q_A$). Frequency response of the nLnH-MetaMater for the **f,** left incidence, **g,** right incidence at the incidence frequencies (1903 and 1919 Hz) corresponding to the two highest transmission asymmetries. **h,** frequency response in the empty impedance tube for the speaker excitation at 1903 and 1919 Hz.

The transmission asymmetry can be understood qualitatively by using a geometric series consideration similar to that of the LnH-MetaMater. The terms that describe the sound-wave-paths for left (Al side) and right (PC side) incidences include "time-reversal" pairs such as $t^{Al}t^{PC}$ and $t^{PC}t^{Al}$, $t^{Al}r^{PC}r^{Al}t^{PC}$ and $t^{PC}r^{Al}r^{PC}t^{Al}$, etc., which regardless of containing the same sequence of scattering events with PC and Al, but in reverse order, are no more equal. Specifically, the transmission $t^{Al/PC}$ and reflection $r^{Al/PC}$ terms are amplitude dependent. As a result, they are affected by the direction of the sound incidence, i.e. $t_{left}^{Al/PC} \neq t_{right}^{Al/PC}$ and $r_{left}^{Al/PC} \neq r_{right}^{Al/PC}$. The exact summation of such nonlinear (amplitude dependent due to incidence direction) series in the case of the nLnH-MetaMater is a difficult task. Nevertheless, one can get a quantitative understanding of the transmission asymmetry of the nLnH-MetaMater by modeling the system as a set of two coupled oscillators.

**Numerical Model.** When the principal and sway modes are well-separated spectrally, the tuning fork system can be modeled as two coupled masses (Fig.5(a)). By adding a semi-infinite lead of identical coupled masses on each side we are able to investigate the scattering properties of the system. In our model, the oscillator that describes the PC fork (blue circle in Fig.5(a)) has nonlinear losses while the other oscillator that describes the Al fork (red circle in Fig.5(a)) is modeled as a linear lossless oscillator. The oscillators are coupled with a spring that has a coupling constant $k$, which models the distance between the forks (the shorter the distance the higher the coupling constant $k$). At its left and right, this system is coupled to two semi-infinite arrays of linear oscillators (black circle in Fig.5(a)) that propagate the sound in free space (Fig.5(a)).

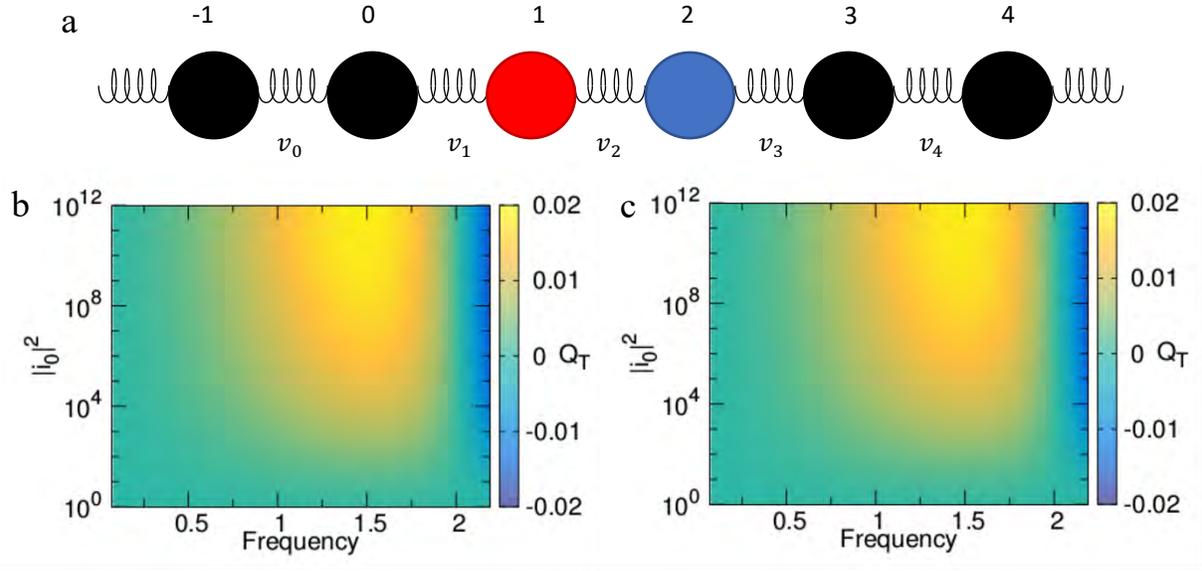

**Figure 5| Numerical model: a,** Illustration of the nLnH-MetaMater model (the top and bottom labels indicate the index, $n$, of each mass and the spring constant, $v$, of each coupling, respectively), **b,** Asymmetry in transmittance $Q_T$ as a function of frequency and incident intensity assuming both lossy and real nonlinearity (here, $a_I = 0.06$, and $a_R = 0.02$), and **c,** Asymmetry in transmittance $Q_T$ as a function of frequency and incident intensity assuming only the lossy nonlinearity (same $a_I$ as in **b,** $a_R = 0$).

The equation of motion for each of the masses shown in Fig.5(a) is given by,

$$m_n\ddot{x}_n = -v_n(x_n - x_{n-1}) - v_{n+1}(x_n - x_{n+1}) - (\mu_n + \tilde{f}_n(x_n))\dot{x}_n + \tilde{g}_n(x_n)x_n \quad (1),$$

where $m_n$ is the $n^{\text{th}}$ mass, $x_n$ is the displacement of the $n^{\text{th}}$ mass, $v_n$ is the spring constant of the spring connecting the $n^{\text{th}}$ mass to the $(n-1)^{\text{th}}$ mass, $\mu_n$ is the linear friction coefficient, and we have assumed that the masses are at a unit distance between one-another at their equilibrium points. The arbitrary functions $\tilde{f}_n(x_n)$ and $\tilde{g}_n(x_n)$ describe the lossy and real nonlinearity of the system respectively. Of these parameters, $\mu_n$, $\tilde{f}_n$, and $\tilde{g}_n$ are zero for $n \neq 1, 2$. For the spring constants and masses, we have,

$$v_n = \begin{cases} v, & n = 2 \\ v_0, & n \neq 2 \end{cases}$$

$$m_n = \begin{cases} m_1, & n = 1 \\ m_2, & n = 2 \\ m_0, & n \neq 1,2 \end{cases} \qquad (2).$$

Dividing Eq. (1) by $m_0$ results in,

$$\frac{m_n}{m_0}\ddot{x}_n = -\kappa_n(x_n - x_{n-1}) - \kappa_{n+1}(x_n - x_{n+1}) - (\gamma_n + f_n(x_n))\dot{x}_n + g_n(x_n)x_n \quad (3),$$

where $\kappa_n \equiv v_n/m_0$, $\gamma_n \equiv \mu_n/m_0$, $f_n \equiv \tilde{f}_n/m_0$, and $g_n \equiv \tilde{g}_n/m_0$.

By assuming harmonic solutions of the form $x_n = a_n e^{-i\omega t}$, we obtain:

$$-\omega^2 \frac{m_n}{m_0} x_n = -\kappa_n(x_n - x_{n-1}) - \kappa_{n+1}(x_n - x_{n+1}) + i\omega(\gamma_n + f_n(x_n))x_n + g_n(x_n)x_n \quad (4).$$

From Eq. (4) and using the lead values for $\kappa_n$, $\gamma_n$, $f_n$, and $g_n$, it can be shown that the leads can support propagating waves with a dispersion relation:

$$\omega = \sqrt{2\kappa_0(1 - \cos k)} \qquad (5)$$

where $k$ is the associated wavenumber of the propagating wave with frequency $\omega$.

The transport properties of the coupled nonlinear oscillators of Eq.(2,4) can be further investigated using the backward transfer map:

$$x_{n-1} = -\frac{\kappa_{n+1}}{\kappa_n}x_{n+1} + \left(1 + \frac{\kappa_{n+1}}{\kappa_n} - \frac{\omega^2}{\kappa_n}\frac{m_n}{m_0} - i\frac{\omega}{\kappa_n}(\gamma_n + f_n(x_n)) - \frac{g_n(x_n)}{\kappa_n}\right)x_n \quad (1),$$

which super-imposed with appropriate scattering boundary conditions can allow for a unique solution of the nonlinear problem. We assume incident and transmitted plane waves of the form:

$$x_n = \begin{cases} i_0 e^{ikn} + r_0 e^{-ikn} & n \leq 1 \\ t_0 e^{ikn} & n \geq 2 \end{cases} \qquad (2),$$

where $i_0$, $r_0$, and $t_0$ are the complex amplitudes of the incident, reflected, and transmitted waves, respectively. Scattering boundary conditions Eq. (7) assume a left incident wave. For a right incident wave one needs to modify Eqs. (7) accordingly.

We can obtain $i_0$ and $r_0$ in terms of $x_1$ and $x_2$ from Eq. (7):

$$i_0 = \frac{x_1 - x_0 e^{-ik}}{e^{ik} - e^{-ik}}, \quad r_0 = \frac{x_1 - x_0 e^{ik}}{e^{-ik} - e^{ik}} \tag{3}$$

These relations, combined with Eq. (5,6), allow us to obtain $i_0$ and $r_0$ in terms of $t_0$ and $k$ by recursively calculating the displacement of each of the masses, and in turn we can compute reflectance and transmittance for the system. This way of calculating the scattering properties of nonlinear systems by fixing the output scattering conditions is necessary because of the multi-stability often present in these problems (D'Ambroise et al., 2012; Hennig and Tsironis, 1999; Lepri and Casati, 2011).

We evaluate the transmittance $T \equiv \left|\frac{t_0}{i_0}\right|^2$ and reflectance $R \equiv \left|\frac{r_0}{i_0}\right|^2$ using the backward map approach. First, we perform the transformation $A_n \equiv \frac{x_n e^{-i2k}}{t_0}$. This allows us to simplify the form of the wave at the output. Assuming for simplicity that $m_0 = 1$, Eq. (6) becomes,

$$A_{n-1} = -\frac{\kappa_{n+1}}{\kappa_n} A_{n+1} + \left(1 + \frac{\kappa_{n+1}}{\kappa_n} - \frac{m_n}{\kappa_n}\omega^2 - i\frac{\omega}{\kappa_n}(\gamma_n + F_n(A_n)) - \frac{G_n(A_n)}{\kappa_n}\right) A_n \tag{4},$$

where $F_n(A_n) \equiv f_n(x_n(A_n))$, $G_n(A_n) \equiv g_n(x_n(A_n))$. From (7), we have $A_3 = e^{ik}$, $A_2 = 1$. By recursively applying Eq. (9), we obtain,

$$A_1 = \delta - \frac{\kappa_0}{\kappa}e^{ik}$$
$$A_0 = -\frac{\kappa}{\kappa_0} + \alpha(\delta - \frac{\kappa_0}{\kappa}e^{ik})$$
$$\delta = 1 + \frac{\kappa_0}{\kappa} - \frac{m_2}{\kappa}\omega^2 - i\frac{\omega}{\kappa}(\gamma_2 + F_2(1)) - \frac{G_2(1)}{\kappa} \tag{5},$$
$$\alpha = 1 + \frac{\kappa}{\kappa_0} - \frac{m_1}{\kappa_0}\omega^2 - i\frac{\omega}{\kappa_0}(\gamma_1 + F_1(\delta - \frac{\kappa_0}{\kappa}e^{ik})) - \frac{G_1(\delta - \frac{\kappa_0}{\kappa}e^{ik})}{\kappa_0}$$

which together with Eqs. (8) allow us to obtain the transmittance and reflectance:

$$T = \left|\frac{t_0}{i_0}\right|^2 = \left|\frac{e^{ik} - e^{-ik}}{\frac{\kappa}{\kappa_0} - (\delta - \frac{\kappa_0}{\kappa}e^{ik})(a - e^{ik})}\right|^2$$
$$R = \left|\frac{r_0}{i_0}\right|^2 = \left|\frac{\frac{\kappa}{\kappa_0} - (\delta - \frac{\kappa_0}{\kappa}e^{ik})(a - e^{-ik})}{\frac{\kappa}{\kappa_0} - (\delta - \frac{\kappa_0}{\kappa}e^{ik})(a - e^{ik})}\right|^2 \tag{6}.$$

As we discussed previously, the above calculations assume left-incident scattering boundary conditions, see Eqs. (7). To evaluate the transmittance and reflectance for a right-incident wave one needs to flip the subscripts of $m$, $\gamma$, $F$, and $G$ that appear in the expressions for $\alpha$ and $\delta$ in Eqs. (10), i.e. $1 \rightarrow 2, 2 \rightarrow 1$, which is equivalent to flipping the sample.

Equations (11) represent the most general solution of the non-linear scattering problem. To make a qualitative comparison with the experimental results we assume specific functions for the nonlinear functions $F_{1,2}$ and $G_{1,2}$ (associated with $f_{1,2}$ and $g_{1,2}$ respectively). From the experimental observations of the resonant behavior of the polycarbonate fork (resonant frequency and FWHM vs. SPL), we obtain the functional form of $g_2$ and $f_2$ vs. displacement respectively, mapping displacement to sound pressure. The functions used are:

$$
\begin{aligned}
f_2(x_2) &= a_I(\log_{10}(1 + |x_2|))^2 \\
g_2(x_2) &= a_R(\log_{10}(1 + |x_2|))^2
\end{aligned}
\implies
\begin{aligned}
F_2(A_2) &= a_I(\log_{10}(1 + |tA_2|))^2 \\
G_2(A_2) &= a_R(\log_{10}(1 + |tA_2|))^2
\end{aligned}
\quad (7),
$$

where $a_I$, $a_R$ are real numbers. Assuming that the nonlinearity of the aluminum fork and its linear losses are negligible, $g_1$, $f_1$ and, $\gamma_1$ are equal to zero. We also adjust $m_1$ so that the resonators 1 and 2 have the same resonant frequency in spite of the losses in resonator 2.

We then obtain the transmittance from the left ($T_L$) and right ($T_R$) for a range of incident frequencies $\omega$ and transmitted wave amplitudes $t_0$ using Eqs. (5,11). Once we have these, we compute the incident wave intensity for a given frequency and transmitted wave amplitude as $|i_0|^2 = \frac{|t_0|^2}{T}$. We have observed multistability (i.e. multiple possible output waves for a single input) in our results for a range of parameters (i.e. $|i_0|^2$ corresponding to more than one values of $T$), mostly around small frequencies and relatively large values of $|i_0|^2$ when $g_2(x_2)$ is large. However, multistabilities are suppressed for the parameter range that is presented below. To verify that the model reproduces the features of the system observed in experiments, we interpolate the results of $T_{L/R}(|i_0|^2, \omega)$ so that we can obtain the degree of asymmetry $Q_T = \frac{T_R - T_L}{T_R + T_L}$. Obviously, $Q_T$ can only be obtained for the parameter range where both $T_{L/R}$ are single-valued functions of the incident intensity.

In Fig.5(b,c), we present our computed $Q_T$ vs. the frequency in the x-axis, which is normalized by $\omega_0 = \sqrt{\frac{\kappa_0}{m_1}}$, and as a function of the incident signal intensity in the y-axis, for two sets of parameters: (a) combination of lossy and real nonlinearities at the resonator modeling the polycarbonate fork, (b) lossy nonlinearity only. The rest of the parameters used are $\kappa = 1.35$, $\kappa_0 = 1$, $m_1 = 1.26$, $m_2 = 1.20$, $\gamma_1 = 0$, $\gamma_2 = 0.5$. In Fig.5(b), we observe that, as the incident intensity increases and the nonlinearity becomes significant, we transition from a region of symmetry in transmission to a region of asymmetry. In the asymmetric region, for small frequencies, $Q_T$ takes small values close to zero. As the frequency increases, $Q_T$ transitions to positive, then back to zero and finally it takes negative values. We also observe a shift to the left of the lobe of positive $Q_T$ in the color-density plot with increasing incident intensity. We notice a very similar behavior in Fig.5(c), where the real nonlinearity is set to be zero in our calculations, with a slight increase in the shifting of the positive $Q_T$ lobe. This effect can be more dramatic for different nonlinear functions. This result suggests that the dominant effect that leads to asymmetric acoustic energy transmission and its switching behavior we observe in the experiment is the presence of lossy nonlinearity.

To understand the source of asymmetry in transmittance, we examine the dependence of $T_{L/R}$ on the nonlinearities. For simplicity, we will assume $\kappa_0 = \kappa$ and $m_1 \approx m_2 = m$. Starting with the expression for transmittance in (11) and using the appropriate expressions from (10), we obtain,

$$T_L = \left| \frac{2 \sin k}{1 - \eta[\eta - i\tilde{\gamma}_2 - i\tilde{F}_2(1) - \tilde{G}_2(1)]} \right|^2$$

$$T_R = \left| \frac{2 \sin k}{1 - \eta[\eta - i\tilde{\gamma}_2 - i\tilde{F}_2(\eta) - \tilde{G}_2(\eta)]} \right|^2$$

(8),

where $\beta \equiv 2 - \frac{m}{\kappa}\omega^2$, $\eta \equiv \beta - e^{ik}$, $\tilde{\gamma}_2 \equiv \gamma_2 \frac{\omega}{\kappa}$, $\tilde{F}_2(x) \equiv F_2(x)\frac{\omega}{\kappa}$, and $\tilde{G}_2(x) \equiv \frac{G_2(x)}{\kappa}$. The discrepancy between these expressions stems from the arguments of the nonlinear functions. This is true for both the real and the imaginary nonlinear terms. If we assume no real nonlinearity (inspired by the fact that Figs.5b and c are similar) and small values of $x_2$ so that $f_2(x_2) \approx a_I |x_2|^2$ and $F_2(A_2) \approx a_I |tA_2|^2$, Eq.(13) can be further reduced to the form

$$T_L = \left| \frac{A}{B + i\Gamma} \right|^2$$

$$T_R = \left| \frac{A}{B + i(\Gamma + \tilde{\Gamma})} \right|^2$$

(9),

with $A \equiv 2 \sin k$, $B \equiv 1 - \eta^2$, $\Gamma \equiv \eta(\tilde{\gamma}_2 + \tilde{F}_2(1))$, $\tilde{\Gamma} \equiv \eta\tilde{F}_2(1)(|\eta|^2 - 1)$. Eqs. (14) demonstrate how the asymmetry arises even in a simple case of lossy non-linearity.

**Discussions.** From the numerical modeling, we have found that the transmittance of an incident sound wave from the right (PC) side of the nonlinear lossy oscillator is $T_{PM} = \left| \frac{A}{B + i(\Gamma + \tilde{\Gamma})} \right|^2$. Similarly, the transmittance of an incident wave from the left (Al) side of the linear lossless oscillator is $T_{MP} = \left| \frac{A}{B + i\Gamma} \right|^2$. The variables A and B contain "linear" information such as the frequency of the incident wave, the dispersion properties of the propagating medium, structural/geometric information of the resonant cavity. The variable $\Gamma$ includes information about the linear and nonlinear losses associated with the lossy oscillator. Typically, $\Gamma$ is an increasing function of the excitation amplitude and is responsible for the nonlinear resonant shift towards lower frequencies that we observe in the transmission spectrum of the nLnH-MetaMater (Fig.5). The term that is responsible for the asymmetry in the transmittance is $\widetilde{\Gamma}$, which is also an increasing function of the excitation amplitude. It incorporates nonlinear (i.e. amplitude-dependent) losses and directionality—it is zero for left-incident waves and different from zero for right-incident waves. In the latter case, it adds to the existing reciprocal losses $\Gamma$, thus 'pushing' the resonant transmittance for the right-incident wave further to lower frequency values. In other words, $T_{PM/MP}(\omega)$ get their corresponding maximum values at different resonant frequencies $\omega = \omega_{PM}^{res} < \omega_{MP}^{res}$ respectively. The separation between $\omega_{PM}^{res}, \ \omega_{MP}^{res}$ becomes larger as the amplitude of the incident wave (and thus $\widetilde{\Gamma}$) increases. Consequently, the transmittance asymmetry $Q_T(\omega_{PM}^{res}) > 0$ as the amplitude of the incident wave increases. This behavior further verifies our experimental results shown in Fig. 4.

We also find from the coupled oscillator model that the asymmetric transport can additionally be affected by the coupling strength between the oscillators. By changing the center-to-center gap, G (Fig.S1b), between the two forks in the nLnH-MetaMater, we demonstrate the effect of coupling strength (Fig.S5, S6). This suggests that the asymmetric energy transport in the nLnH-MetaMater can be controlled not only as a function of the incident amplitude, but also as a function of the gap between the constituent resonant elements. Interestingly, for a gap G=38 mm, the reflection is completely suppressed (R_MP=0) for left incidence, leading to a unidirectional extraordinary absorber with absorbance greater than 80% (Fig.S6 g,h,i).

Transmission asymmetry is forbidden by the reciprocity theorem in the case of linear time-reversal symmetric systems, consistent with the symmetric transmittance in LnH-MetaMater that we discussed initially. In contrast, asymmetric acoustic transport can be realized in systems with conservative nonlinearities and in the absence of mirror symmetry (Boechler et al., 2011; Lepri and Casati, 2011; Lepri and Pikovsky, 2014; Liang et al., 2009). In such cases, however, the outgoing signals at the fundamental frequency are noisy because of high insertion losses due to high impedance mismatch (Lepri and Casati, 2011; Liang et al., 2009, 2010; Maznev et al., 2013). Instead, one may utilize the existence of large asymmetric frequency conversion for efficiently transmitting power from one side to the other. The obvious drawback of this approach is that the harvested signal is at different frequencies than the input signal. Use of amplifiers can overcome insertion losses (Popa and Cummer, 2014), which, however, makes the structure bulky while consuming external power. Our compact (thickness ~subwavelength) nLnH-MetaMater overcomes both of those challenges, and provides asymmetric transmission in the same frequency as the incident wave, while consuming no power. Designing a nLnH-MetaMater using different lossy materials such as polymers, domain-switching ferroelectric materials, and carbon nanotube foams where the loss can be controlled actively will enable the creation of novel devices that can control the sound propagation in desired ways for various applications. Acoustic switches, diodes, and unidirectional perfect absorbers are some of the many potential devices that can be developed using nLnH-MetaMater.

**Methods**

Details of the experimental methods, sample fabrication, and additional data can be found in Supplementary Information.

**Additional Information**

Supplementary information is available in the online version of the paper.

**Acknowledgements**

We acknowledge the financial support from the Air Force Office of Scientific Research via a Multi-University Research Initiative (Contract no. FA9550-14-1-0037). We also acknowledge

David Smith, Michael Bonar, and Gary L. Woods of Rice University, and Miguel Moleron of ETH Zurich for their support during the acoustic testing apparatus development.

**Competing Financial Interests**

Authors declare no competing financial interests.

**References**

Åbom, M. (**1991**). "Measurement of the scattering-matrix of acoustical two-ports," Mech. Syst. Signal Process., **5**, 89–104. doi:10.1016/0888-3270(91)90017-Y

Boechler, N., Theocharis, G., and Daraio, C. (**2011**). "Bifurcation-based acoustic switching and rectification," Nat. Mater. doi:10.1038/NMAT3072

D'Ambroise, J., Kevrekidis, P. G., and Lepri, S. (**2012**). "Asymmetric wave propagation through nonlinear PT-symmetric oligomers," J. Phys. A Math. Theor. doi:10.1088/1751-8113/45/44/444012

Deymier, P. A. (**2013**). *Acoustic Metamaterials and Phononic Crystals*, (P. A. Deymier, Ed.) Springer Series in Solid-State Sciences, Springer Berlin Heidelberg, Berlin, Heidelberg, Vol. 173. doi:10.1007/978-3-642-31232-8

El-Ganainy, R., Makris, K. G., Christodoulides, D. N., and Musslimani, Z. H. (**2007**). "Theory of coupled optical PT-symmetric structures," Opt. Lett., **32**, 2632. doi:10.1364/OL.32.002632

El-Ganainy, R., Makris, K., Khajavikhan, M., Musslimani, Z., Rotter, S., and Christodoulides, D. (**2018**). "Non-Hermitian physics and PT symmetry," Nat. Phys., doi: 10.1038/nphys4323. doi:10.1038/nphys4323

Eleftheriades, G. V., and Balmain, K. G. (**2005**). *Negative-Refraction Metamaterials*, (G. V. Eleftheriades and K. G. Balmain, Eds.) John Wiley & Sons, Inc., Hoboken, NJ, USA. doi:10.1002/0471744751

Fleury, R., Sounas, D., and Alù, A. (**2015**). "An invisible acoustic sensor based on parity-time symmetry," Nat. Commun., **6**, 5905. doi:10.1038/ncomms6905

Fleury, R., Sounas, D., Haberman, M. R., and Alù, A. (**2015**). "Nonreciprocal Acoustics," Acoust. Today, **11**, 14–21.

Haberman, M. R., and Guild, M. D. (**2016**). "Acoustic metamaterials," Phys. Today, **69**, 42–48. doi:10.1063/PT.3.3198

Hennig, D., and Tsironis, G. P. (**1999**). "Wave transmission in nonlinear lattices," Phys. Rep., **307**, 333–432. doi: 10.1016/S0370-1573(98)00025-8

Hodaei, H., Hassan, A. U., Wittek, S., Garcia-Gracia, H., El-Ganainy, R., Christodoulides, D. N., and Khajavikhan, M. (**2017**). "Enhanced sensitivity at higher-order exceptional points," Nature, **548**, 187–191. doi:10.1038/nature23280

*ISO 10534-2:1998(E) Acoustics--Determination of sound absorption coefficient and impedance in impedance tubes-- Part 2: Transfer-function method*, (**1998**). The International Organization for Standardization, First edit.

Joannopoulos, J. D., Johnson, S. G., Winn, J. N., and Meade, R. D. (**2008**). *Photonic Crystals: Modeling the flow of light*, Princeton University Presss, Princeton, NJ, 2nd Editio.

Konotop, V. V, Yang, J., and Zezyulin, D. A. (**2016**). "Nonlinear waves in PT-symmetric systems," Rev. Mod. Phys., **88**, 35002. doi:10.1103/RevModPhys.88.035002

Kottos, T., and Aceves, A. B. (**2016**). "Synthetic Structures with Parity-Time Symmetry," In O. Shulika and I. Sukhoivanov (Eds.), Contemp. Optoelectron. Mater. Metamaterials Device Appl., Springer Netherlands, Dordrecht, pp. 147–162. doi:10.1007/978-94-017-7315-7_9

Lakes, R. S. (**2001**). "Extreme Damping in Composite Materials with a Negative Stiffness Phase," Phys. Rev. Lett., **86**, 2897–2900. doi:10.1103/PhysRevLett.86.2897

Lee, J.-H., Singer, J. P., and Thomas, E. L. (**2012**). "Micro-/nanostructured mechanical metamaterials," Adv. Mater., **24**, 4782–810. doi:10.1002/adma.201201644

Lepri, S., and Casati, G. (**2011**). "Asymmetric Wave Propagation in Nonlinear Systems," Phys. Rev. Lett., **106**, 164101. doi:10.1103/PhysRevLett.106.164101

Lepri, S., and Pikovsky, A. (**2014**). "Nonreciprocal wave scattering on nonlinear string-coupled oscillators," Chaos. doi:10.1063/1.4899205

Liang, B., Guo, X. S., Tu, J., Zhang, D., and Cheng, J. C. (**2010**). "An acoustic rectifier," Nat Mater, **9**, 989–992.

Liang, B., Yuan, B., and Cheng, J. (**2009**). "Acoustic Diode: Rectification of Acoustic Energy Flux in One-Dimensional Systems," Phys. Rev. Lett., **103**, 104301. doi:10.1103/PhysRevLett.103.104301

Makris, K. G., El-Ganainy, R., Christodoulides, D. N., and Musslimani, Z. H. (**2008**). "Beam Dynamics in PT Symmetric Optical Lattices," Phys. Rev. Lett., **100**, 103904. doi:10.1103/PhysRevLett.100.103904

Maznev, A. A., Every, A. G., and Wright, O. B. (**2013**). "Reciprocity in reflection and transmission: What is a 'phonon diode'?," Wave Motion, **50**, 776–784. doi:10.1016/j.wavemoti.2013.02.006

Merkel, A., Theocharis, G., Richoux, O., Romero-García, V., and Pagneux, V. (**2015**). "Control of acoustic absorption in one-dimensional scattering by resonant scatterers," Appl. Phys. Lett. doi:10.1063/1.4938121

Milton, G. W., Briane, M., and Willis, J. R. (**2006**). "On cloaking for elasticity and physical equations with a transformation invariant form," New J. Phys., **8**, 248.

Molerón, M., Serra-Garcia, M., and Daraio, C. (**2015**). "Visco-thermal effects in acoustic

metamaterials: from total transmission to total reflection and high absorption," New J. Phys., **18**, ArXiv: 1511.05594. doi:10.1088/1367-2630/18/3/033003

Musslimani, Z. H., Makris, K. G., El-Ganainy, R., and Christodoulides, D. N. (**2008**). "Optical solitons in PT symmetric periodic potentials," Phys. Rev. Lett., **100**, 030402. doi:10.1103/PhysRevLett.100.030402

Nassar, H., Xu, X. C., Norris, A. N., and Huang, G. L. (**2017**). "Modulated phononic crystals: Non-reciprocal wave propagation and Willis materials," J. Mech. Phys. Solids, **101**, 10–29. doi:10.1016/j.jmps.2017.01.010

Popa, B.-I., and Cummer, S. a (**2014**). "Non-reciprocal and highly nonlinear active acoustic metamaterials," Nat. Commun., **5**, 3398. doi:10.1038/ncomms4398

Rossing, T. D. (**1992**). "On the acoustics of tuning forks," Am. J. Phys. doi:10.1119/1.17116

Saleh, B. E. A., and Teich, M. C. (**1991**). *Fundamentals of Photonics*, Wiley Series in Pure and Applied Optics, John Wiley & Sons, Inc., New York, USA. doi:10.1002/0471213748

Shi, C., Dubois, M., Chen, Y., Cheng, L., Ramezani, H., Wang, Y., and Zhang, X. (**2016**). "Accessing the exceptional points of parity-time symmetric acoustics," Nat. Commun., **7**, 1–5. doi:10.1038/ncomms11110

Song, J. Z., Bai, P., Hang, Z. H., and Lai, Y. (**2014**). "Acoustic coherent perfect absorbers," New J. Phys. doi:10.1088/1367-2630/16/3/033026

Suchkov, S. V, Sukhorukov, A. A., Huang, J., Dmitriev, S. V, Lee, C., and Kivshar, Y. S. (**2016**). "Nonlinear switching and solitons in PT-symmetric photonic systems," Laser Photon. Rev., **10**, 177–213. doi:10.1002/lpor.201500227

Werner, D. H., and Kwon, D.-H. (**2014**). *Transformation Electromagnetics and Metamaterials*, (D. H. Werner and D.-H. Kwon, Eds.) Springer London, London. doi:10.1007/978-1-4471-

4996-5

Zalevsky, Z., and Mendlovic, D. (**2004**). *Optical Superresolution*, Springer Series in Optical

Sciences, Springer New York, New York, Vol. 91. doi:10.1007/978-0-387-34715-8

Zhu, X., Ramezani, H., Shi, C., Zhu, J., and Zhang, X. (**2014**). "PT Symmetric Acoustics," Phys.

Rev. X, **4**, 031042. doi:10.1103/PhysRevX.4.031042

Ramathasan Thevamaran[1,=,*], Richard Massey Branscomb[2,=], Eleana Makri[3], Paul Anzel[4],

Demetrios Christodoulides[5], Tsampikos Kottos[3], Edwin L. Thomas[2,*]

[1]*Department of Engineering Physics, University of Wisconsin-Madison, Madison, WI 53706.*

[2]*Department of Materials Science and NanoEngineering, Rice University, Houston, TX 77005.*

[3]*Department of Physics, Wesleyan University, Middletown, CT 06459.*

[4]*HEB, San Antonio, TX 78204.*

[5]*Center for Research and Education in Optics and Lasers (CREOL), College of Optics and Photonics,*

*University of Central Florida, Orlando, FL 32816.*

[*]Corresponding Authors: RT (thevamaran@wisc.edu); ELT (elt@rice.edu)

[=]These authors contributed equally to this work.

**Methods**
**Figures S1-S7**

**Methods**

    **Experimental technique.** The two-port acoustic testing setup consists of a custom-built aluminum impedance tube that has 6.35 mm wall thickness and 25.4 mm x 10 mm rectangular cross-section that matches the configuration requirements for the test sample. It was designed according to ISO 10534-2:1998(E) to allow only the propagation of plane wave modes up to a cutoff frequency of ~6750 Hz. A tweeter (CAT 308) with a power amplifier (Behringer A500) was used to generate the sound waves at one end of the impedance tube. Transition from circular cross-section of the speaker to the rectangular cross-section of the impedance tube was provided by a 3D-printed polycarbonate cone. The other end of the tube was filled with a polyurethane absorbent foam for ~200 mm length to minimize back-reflections. Four pressure-field impedance tube microphones with preamplifiers (B&K 4187), placed as a pair in each port were used to simultaneously measure the propagating waves. A signal conditioning microphone power supply (B&K 2829) was used to power the microphones as well as to condition the measured signal. A digital data acquisition (DAQ) system (NI CDAQ-9174 with a NI 9263 output module and two NI 9215 input modules) was used for signal processing. All the instruments were controlled by a MATLAB program, and a dual-phase lock-in amplification system was implemented for phase-sensitive measurements of the acoustic waves.

    The lock-in amplification system simultaneously generates a cosine and sine waves by NI 9263 DAQ, and simultaneously measures the four microphone signals along with those two cosine and sine waves by NI 9215 DAQs. The voltage amplitude ($V_R$) and phase ($\Phi$) of the signal from each microphone ($V_{sig}$) were calculated by $V_R = \sqrt{V_x{}^2 + V_y{}^2}$ and $\Phi = \tan^{-1}\left(\frac{V_y}{V_x}\right)$, where $V_x = \langle V_{sig}\cos(\omega t + \varphi)\rangle$ and $V_y = \langle V_{sig}\sin(\omega t + \varphi)\rangle$, and then converted into pressure using the calibration factors of each microphone. At each frequency of the frequency sweep, 700 cycles were generated and middle 50% of the measured signals were used for analysis, neglecting the initial and final 25% for any unsteady responses. From these signals, the incident and reflected signals in each port are calculated as described below:

The measured pressures by microphones 1-4 (<span style="color:orange">Fig.S1</span>) are given by,

$$P_1 = P_i^a\, e^{ikx_1} + P_r^a e^{-ikx_1} \qquad (1)$$
$$P_2 = P_i^a e^{ikx_2} + P_r^a e^{-ikx_2} \qquad (2)$$
$$P_3 = P_i^b e^{ikx_3} + P_r^b e^{-ikx_3} \qquad (3)$$
$$P_4 = P_i^b e^{ikx_4} + P_r^b e^{-ikx_4} \qquad (4),$$

where $P_i^a$ and $P_i^b$ are the forward traveling (incidence) and $P_r^a$ and $P_r^b$ are the backward traveling (reflection) pressure waves in port a and b respectively, and $x_1, x_2, x_3, x_4$ are the distances to microphones 1-4 from an arbitrary reference.

From equations (1) and (2), $P_i^a$ and $P_r^a$ in port-a can be calculated as,

$$P_i^a = \frac{P_1 e^{-ikx_2} - P_2 e^{-ikx_1}}{e^{-ik(x_2-x_1)} - e^{ik(x_2-x_1)}} \qquad (5)$$
$$P_r^a = \frac{-(P_1 e^{ikx_2} - P_2 e^{ikx_1})}{e^{-ik(x_2-x_1)} - e^{ik(x_2-x_1)}} \qquad (6)$$

Similarly, in port-b, the transmitted pressure can be calculated as,

$$P_i^b = \frac{P_3 e^{-ikx_4} - P_4 e^{-ikx_3}}{e^{-ik(x_4-x_3)} - e^{ik(x_4-x_3)}} \qquad (7)$$

$$P_r^b = \frac{-(P_3 e^{ikx_4} - P_4 e^{ikx_3})}{e^{-ik(x_4-x_3)} - e^{ik(x_4-x_3)}} \qquad (8) \; (P_r^b \approx 0 \text{ due to foam absorption})$$

Phase-shifting these quantities to the sample's front $(x_f)$ and back $(x_b)$ boundaries results in,

$$P_i = P_i^a \big|_{x=x_f} = \frac{P_1 e^{-ik(x_2-x_1)} - P_2}{e^{-ik(x_2-x_1)} - e^{ik(x_2-x_1)}} \; e^{ik(x_f-x_1)} \qquad (9)$$

$$P_r = P_r^a \big|_{x=x_f} = \frac{-(P_1 e^{ik(x_2-x_1)} - P_2)}{e^{-ik(x_2-x_1)} - e^{ik(x_2-x_1)}} \; e^{-ik(x_f-x_1)} \qquad (10)$$

$$P_t = P_i^b \big|_{x=x_b} = \frac{P_3 e^{-ik(x_4-x_3)} - P_4}{e^{-ik(x_4-x_3)} - e^{ik(x_4-x_3)}} \; e^{-ik(x_3-x_b)} \qquad (11)$$

From these quantities, the reflectance, $R$, the transmittance, $T$, and the absorbance, $A$, are calculated as,

$$R = \frac{P_r P_r^*}{P_i P_i^*} \qquad (12)$$

$$T = \frac{P_t P_t^*}{P_i P_i^*} \qquad (13)$$

$$A = 1 - (R + T) \qquad (14)$$

Here the $P^*$ is the complex conjugate of the pressure $P$, and $k = \frac{2\pi f}{c} - i\delta$ with frequency, $f$ (in Hz), sound speed in air, $c = 346$ m/s, and visco-thermal losses, $\delta$.

We measured the $\delta$ from a similar frequency sweep performed in an empty tube without any sample inside. We obtain the loss factor in units of Nepers/m by calculating the natural logarithm of the transmission per distance given by,

$$\delta = \ln \left( \frac{P_{1f} e^{2\pi i f(x_3-x_1)/c}}{P_{3f}} \right) \Big/ s$$

Here $P_{1f}$ and $P_{3f}$ represent the complex pressure of the forward traveling wave at microphones one and three respectively, and $s$ represents the distance between the microphones one and three. This measurement was repeated and averaged for four different values of $s$.

**Note:** We observed that the reflection at the resonance of the tuning forks was damping the speaker, which can cause up to fifty percent variation in incidence over the whole frequency range. In order to avoid this variation, our instrument control program verifies if the incident sound pressure level (SPL) is within ±0.5 dB of the intended amplitude at each frequency step. If the incident amplitude is outside the above acceptable range, the program re-executes the frequency step by slightly increasing or decreasing the amplitude until the incidence is within the acceptable range. This allowed us to maintain nearly constant incidence throughout all the frequencies tested.

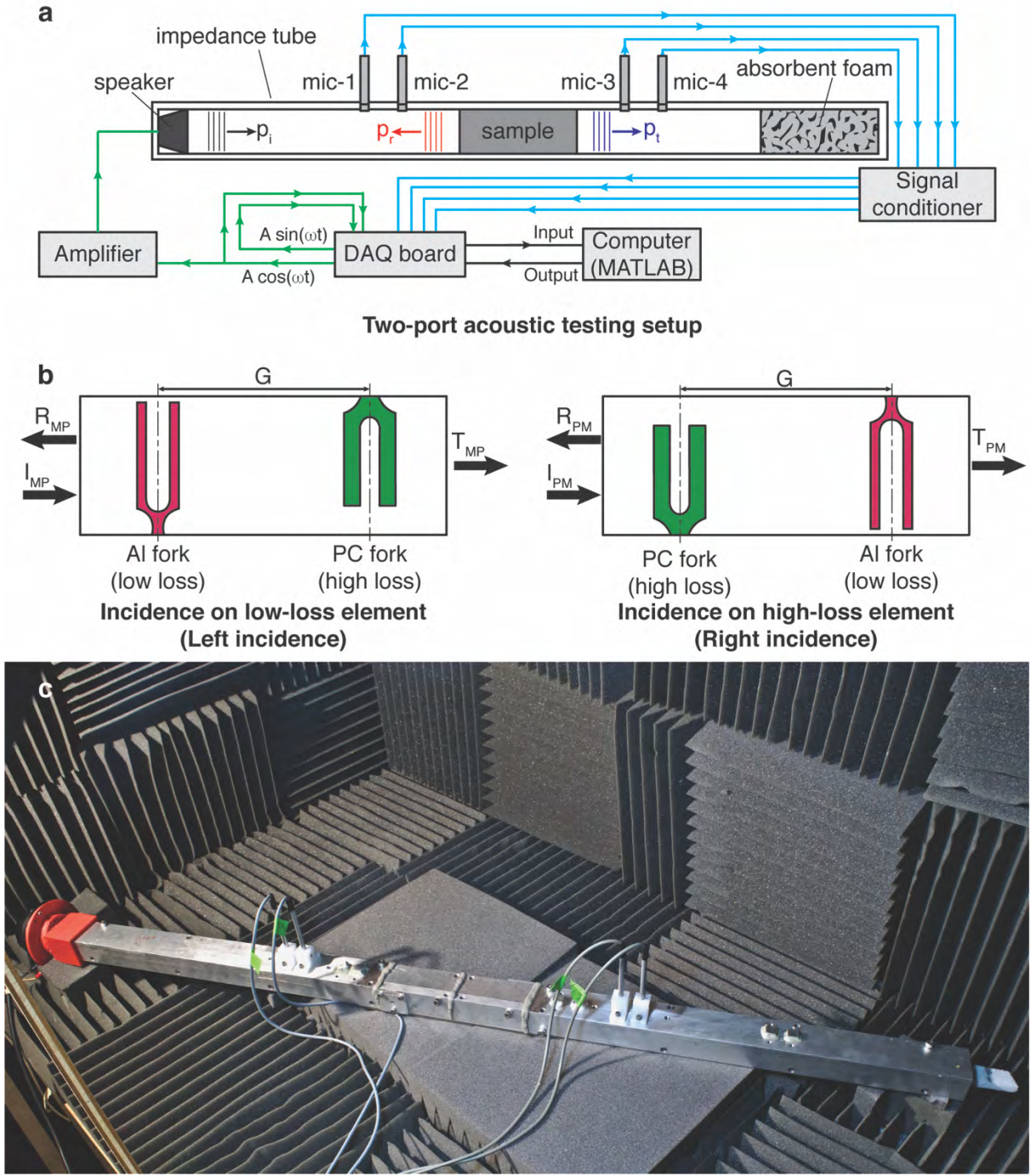

**Figure S1| Illustration and a photograph of the experimental apparatus. a,** schematic illustration of the two-port acoustic testing setup for phase-sensitive measurement of the incidence, reflection, and transmission. **b,** the two sample configurations for investigating directional wave propagation characteristics in the metamaterials; the subscript MP of I, R, and T denotes that the wave is first incident on the metal fork (left incidence), and the subscript PM denotes the wave is first incident on the polymer fork (right incidence). **c,** a photograph of the two-port acoustic testing apparatus in an anechoic chamber.

**Metamaterial Samples.** We fabricated two metamaterials—LnH-MetaMater and nLnH-MetaMater—to investigate the asymmetric acoustic energy transport. The LnH-MetaMater was constructed by machining an aluminum fork and a polycarbonate fork that are matched at their principal mode of resonance. The nLnH-MetaMater was constructed by fabricating an aluminum fork and a polycarbonate fork that are approximately matched at their sway mode of resonance. It should be noted that since the sway mode is highly nonlinear and depends on the excitation amplitude, different initial mismatch in resonance frequencies between the two forks will result in different behavior. The forks arranged as shown in Fig.S1 were firmly mounted using screws on to the walls of a tube section that has the cross-section equal to that of the impedance tube. The gap between the two forks was initially set at 157 mm (~ λ) for both LnH-MetaMater and nLnH-MetaMater. In the case of nLnH-MetaMater, it has been varied to 98 mm and 38 mm to investigate the effects of coupling via air between the forks.

**Supplementary results.**

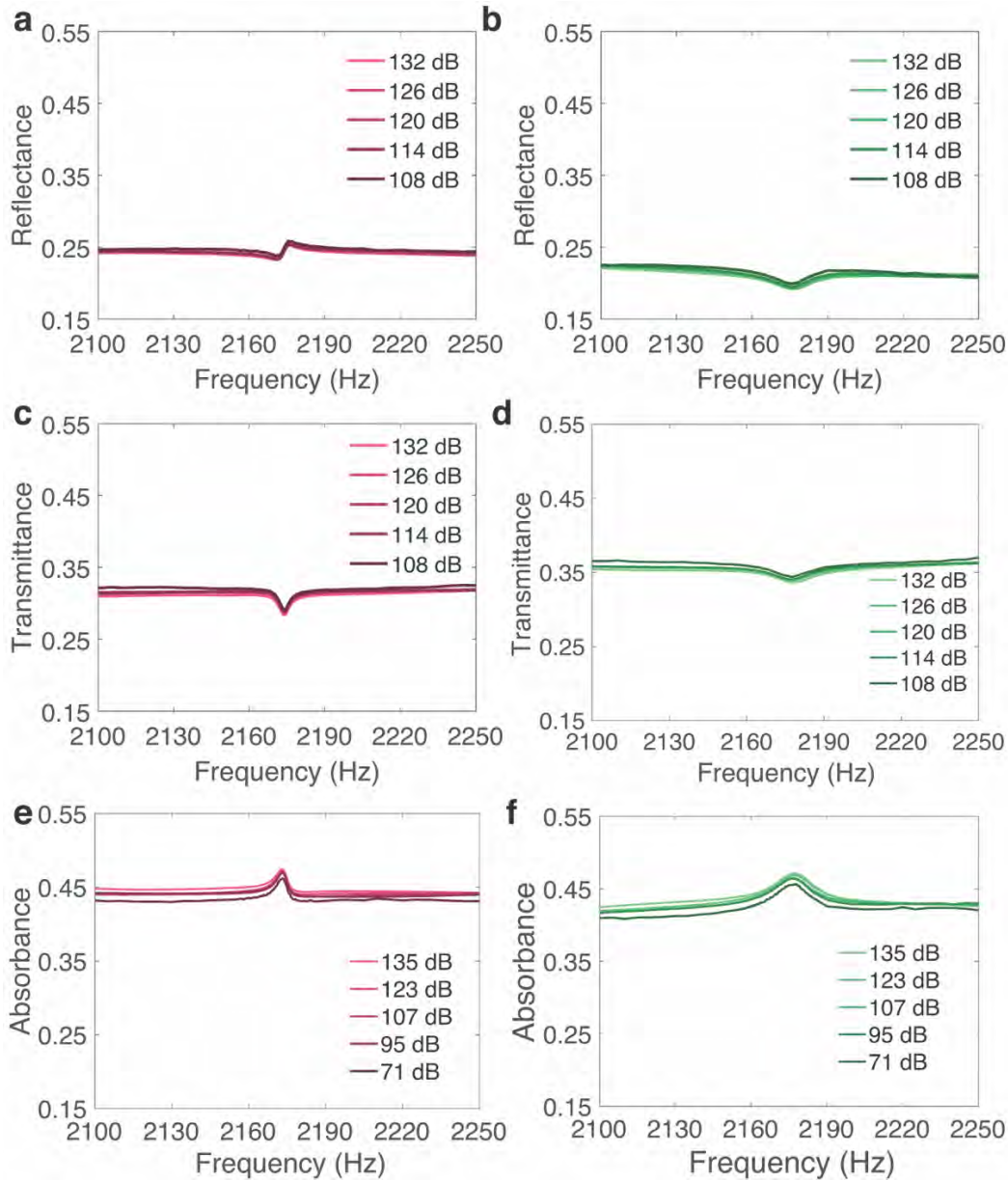

**Figure S2| Resonance of individual forks at principal mode. a,b,** reflectance, **c,d,** transmittance, and **e,f,** absorbance of aluminum (left column, red) and polycarbonate forks (right column, green) at the principal mode of resonance (~2175 Hz) exhibiting an excitation-amplitude-independent linear behavior.

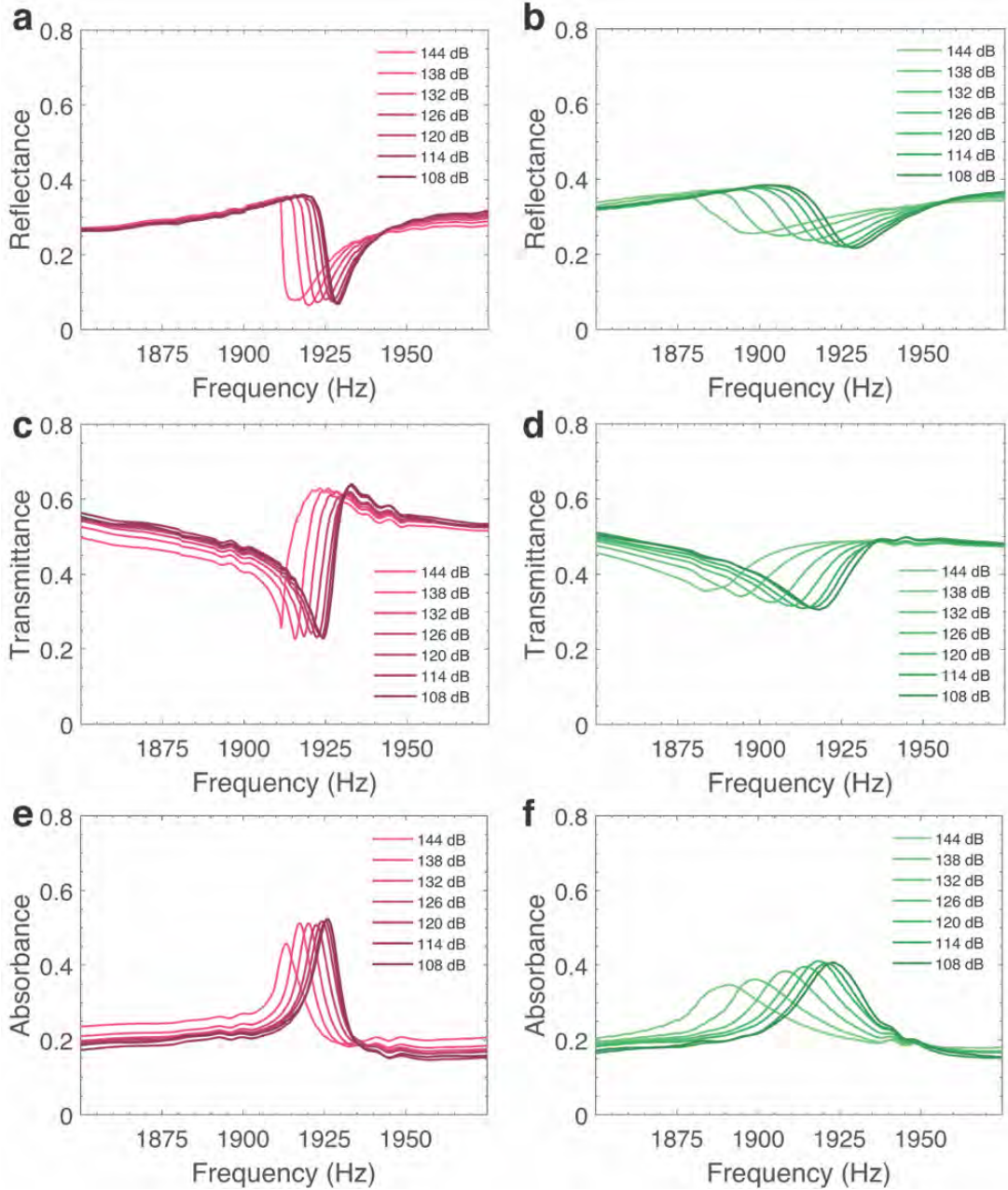

**Figure S3| Resonance of individual forks at sway mode. a,b,** reflectance, **c,d,** transmittance, and **e,f,** absorbance of aluminum (left column, red) and polycarbonate forks (right column, green) at the sway mode of resonance (~1925 Hz) exhibiting an excitation amplitude dependent nonlinear behavior.

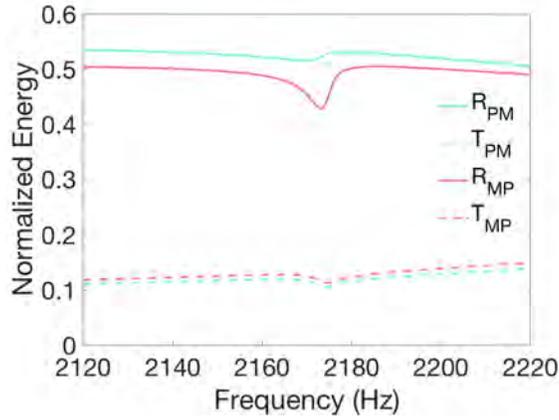

**Figure S4| Transmission and reflection response of the LnH-MetaMater.**

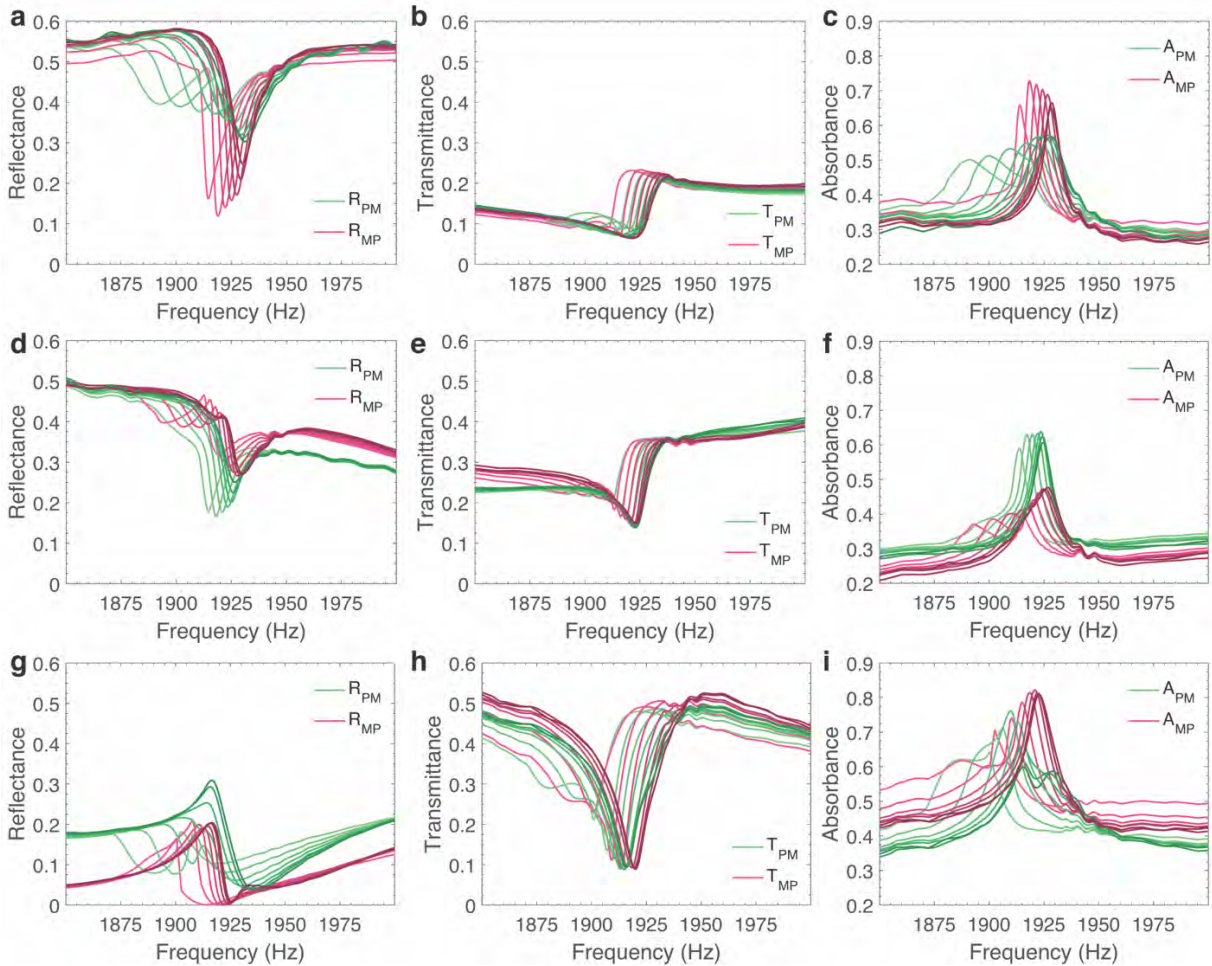

**Figure S5| Asymmetric reflectance, transmittance, and absorbance in nLnH-MetaMater.**
**a,b,c,** Reflectance, transmittance, and absorbance for G=157 mm. **d,e,f,** for G=98 mm. **g,h,i,** for G=38 mm.

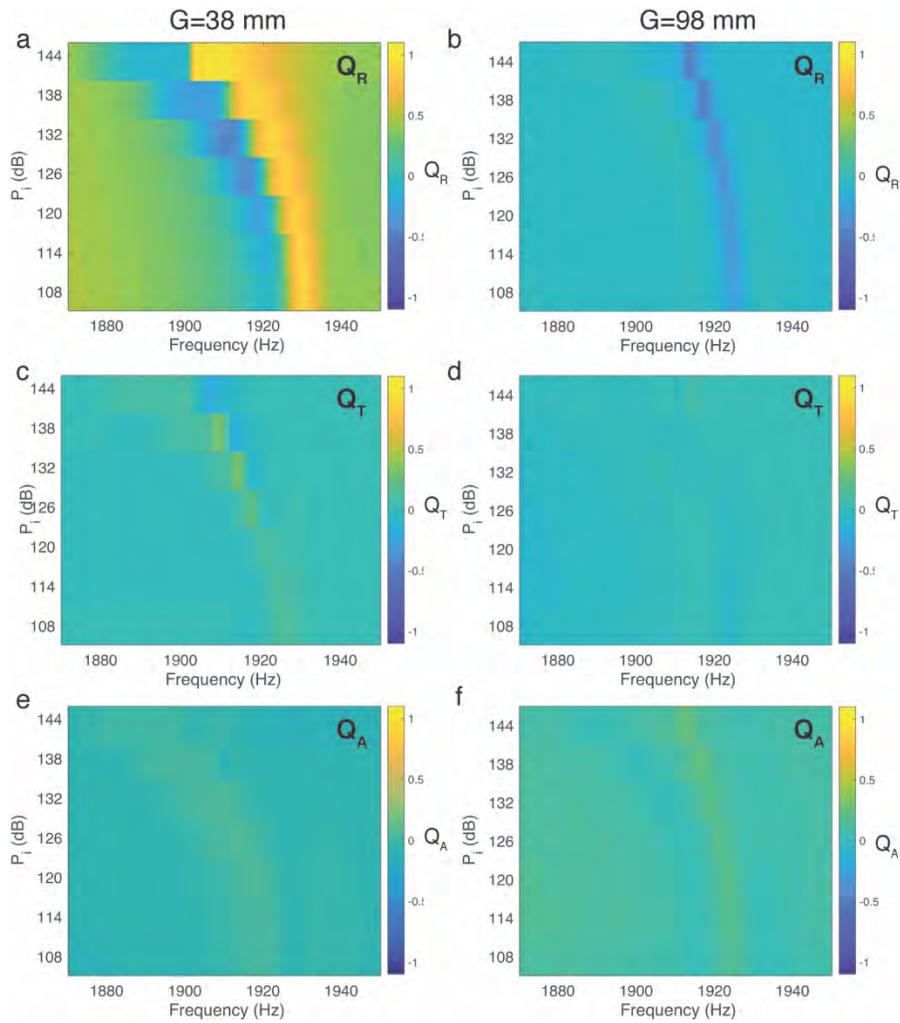

**Figure S6| Effect of the air coupling between the forks on the asymmetric reflectance and transmittance in nLnH-MetaMater.** Asymmetry in **a,b,** reflectance, **c,d,** transmittance, and **e,f,** absorbance for G=38 mm, and G=98 mm, respectively.

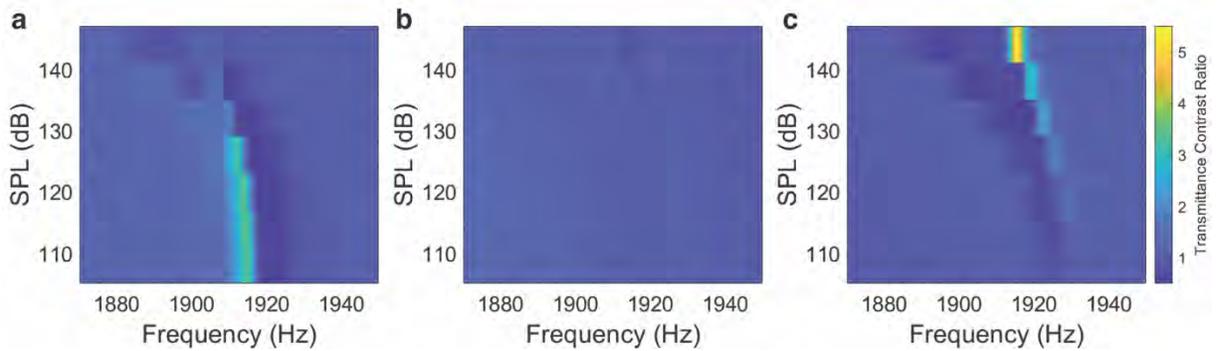

**Figure S7| Contrast ratio (T$_{MP}$/T$_{PM}$) of transmittance asymmetry in nLnH-MetaMater.** Asymmetry in transmittance for **a,** G=38 mm, **b,** G=98 mm, and **c,** G=157 mm respectively.